\documentclass[runningheads]{llncs}

\AtBeginDocument{%
  \providecommand\BibTeX{{%
    \normalfont B\kern-0.5em{\scshape i\kern-0.25em b}\kern-0.8em\TeX}}}
\usepackage{siunitx}
\usepackage{amsmath,amsfonts}
\usepackage{algorithm}
\usepackage{xcolor}
\usepackage{moreverb}
\usepackage{tabularx}
\usepackage{threeparttable} 
\usepackage{multirow} 
\usepackage{rotating} 
\usepackage{bigstrut} 
\usepackage[utf8]{inputenc}
\usepackage{paralist}
\usepackage{lipsum}
\usepackage{soul}
\usepackage[noend]{algpseudocode}
\usepackage{subfig} 
\usepackage{caption}
\usepackage{lipsum}
\usepackage{indentfirst}
\usepackage[labelsep=period]{caption}

\usepackage{todonotes}
\newboolean{showcomments}
\setboolean{showcomments}{true}         

\ifthenelse{\boolean{showcomments}}
{\newcommand{\nb}[2]{
		\fbox{\bfseries\sffamily\scriptsize#1}
		{\sf\small$\blacktriangleright$\textit{\textcolor{red}{#2}}$\blacktriangleleft$}
	}
}
{\newcommand{\nb}[2]{}
	
}

\begin{document}

\title{\textbf{Energy Cyber Attacks to Smart Healthcare Devices: A Testbed}}
\author{Zainab Alwaisi\inst{1}
\and
Simone Soderi\inst{1,2}
\and
Rocco De Nicola\inst{1,2}
}
\authorrunning{Z. Alwaisi et al.}

\institute{IMT School for Advanced Studies, Lucca, Italy\\
\email{\{zainab.alwaisi, simone.soderi, rocco.denicola\}@imtlucca.it} \and
CINI Cybersecurity Laboratory, Roma, Italy 
}

\maketitle              

\begin{abstract}

The Internet of Things (IoT) has emerged as a subject of intense interest among the research and industrial community as it has significantly impacted human life. The rapid growth of
IoT technology has revolutionized human life by inaugurating the concept of smart healthcare, smart devices, smart city, and smart grid. The security of IoT devices has become a serious concern, especially in the healthcare domain, where recent attacks exposed damaging IoT security vulnerabilities. In addition, in IoT networks where the connected devices are vulnerable to attacks such as attacks that affect the resource constraints of healthcare devices, e.g., energy consumption attacks. 
Therefore, this paper defines the impact of Distributed Denial of Service (DDoS) and Fake Access Points (F-APs) attacks on WiFi smart healthcare devices and 
investigates in detail how these attacks can be deployed toward victim devices and Access Points (APs). Our work focuses on IoT devices' connectivity and energy consumption when under attack. The main key findings of this paper are as follows: \begin{inparaenum}[(i)] \item the minimum and maximum attack rate of DDoS attacks that cause service disruptions on the victim side, and \item the minimum-the higher effect of energy-consumption Distributed Denial of Service (EC-DDoS) and F-APs attacks on the energy consumption of the smart healthcare devices\end{inparaenum}. Our study reveals the communication protocols, attack rates, payload sizes, and victim devices' ports state as the vital factors in determining the energy consumption of victim devices. These findings facilitate a thorough understanding of IoT devices' potential vulnerabilities within a smart healthcare environment and pave solid foundations for future studies on defense solutions.

\end{abstract}
\newpage
\noindent\rule{8.4cm}{1pt}\\
Kindly reference this version of the paper:\\

Alwaisi, Z., Soderi, S., Nicola, R.D. (2023). Energy Cyber Attacks to Smart Healthcare Devices: A Testbed. In: Chen, Y., Yao, D., Nakano, T. (eds) Bio-inspired Information and Communications Technologies. BICT 2023. Lecture Notes of the Institute for Computer Sciences, Social Informatics and Telecommunications Engineering, vol 512. Springer, Cham. https://doi.org/10.1007/978-3-031-43135-7\_24
\\

You can reference the following BibTeX entry:
\begin{verbatim}
@InProceedings{10.1007/978-3-031-43135-7_24,
author="Alwaisi, Zainab
and Soderi, Simone
and Nicola, Rocco De",
editor="Chen, Yifan
and Yao, Dezhong
and Nakano, Tadashi",
title="Energy Cyber Attacks to Smart Healthcare Devices: A Testbed",
booktitle="Bio-inspired Information and Communications Technologies",
year="2023",
publisher="Springer Nature Switzerland",
address="Cham",
pages="246--265",
isbn="978-3-031-43135-7",
doi="10.1007/978-3-031-43135-7_24"
}





\end{verbatim}
\noindent\rule{8.4cm}{1pt}
\section{Introduction}

The Internet of Things (IoT) refers to real-world objects having communicative and cognitive capabilities using smart devices. IoT is a tremendous communication paradigm where many heterogeneous devices will connect and talk to each other. These communication devices will play an essential role in the life of human beings, e.g., healthcare, transportation, and others. IoT is creating a revolutionary impact in technology and people's social life. Over time, IoT devices are overgrowing. 
According to recent reports from Cisco, the number of connected smart devices over the Internet will escalate to $29.3$ billion networked devices by 2023~\cite{Cisco2021}.
As the number and heterogeneity of smart devices are accelerating rapidly, it is becoming challenging to maintain the security of these devices~\cite{Shouran2019}.
IoT footprints have been identified in various domains such as manufacturing, agriculture, transportation, electric grid, and healthcare~\cite{He2018}. In an IoT-based healthcare system, security is the primary concern as the data is directly related to human beings~\cite{hireche2022}. An Intensive Care Unit (ICU) is a hospital's special and critically operational department where specialized treatment is given to patients requiring critical medical care. Usually, patients who are acutely unwell or injured severely and require continuous medical care are admitted to the ICU. The equipment and devices in the ICU play a vital role in keeping the patient alive and healthy. In such a scenario, any communication breakdown due to a cybersecurity breach may cause severe effects on a patient's life and even death in some instances~\cite{morgan2021}. 

Moreover, smart healthcare devices typically interact through different wireless communication protocols that allow adversaries to perform different attack types. For example, eavesdropping, creating Fake Access Points (F-APs),
Distributed Denial-of-Service (DDoS), and energy-consumption Distributed-Denial-of-Service (EC-DDoS)~\cite{shahid2022}.
Attackers use DDoS attacks to launch malicious traffic to damage target smart healthcare devices by affecting their resources and disconnecting them from the legitimate AP. 
EC-DDoS attacks lead to increasing the target's energy consumption to destroy it by sending malicious traffic. 
F-APs attacks force smart healthcare devices to connect to an alternative AP, monitor the transferred packets, and launch malicious attacks to consume more energy. 
Typically, most IoT devices have limited processing capabilities, and applying advanced security techniques to each device is  challenging~\cite{Alladi2020}. Using them in the smart healthcare system may also give unauthorized access to cybercriminals to monitor patients' private data and exploit sensitive information or send attacks to consume more energy and destroy smart devices~\cite{YAQOOB2017}. 

Existing studies have performed general static and dynamic analyses and defenses against DDoS and F-AP attacks. However, most of the existing dynamic analyses are conducted in virtual environments, which makes it challenging to accurately measure the compromised devices' resource consumption, especially energy consumption. To address this issue, in this work, we perform our experiments in a controlled environment using real-world devices. With a cost-efficient experimental setup, we can collect sufficient data on the impact of attacks on resource-constrained IoT healthcare devices. 

In this paper, we study the effect of DDoS, EC-DDoS, and F-APs attacks on WiFi connectivity and energy consumption of wireless healthcare devices. The results show the significant damage that could be caused by these attacks and draw attention to the urgent need for effective defense solutions. This paper can be used as a framework to test smart healthcare devices' security and create security standards for robust, predictable, and tamper-free operations.

\subsection{Motivation and Contribution}    
\label{SUBSEC:MOTIVATION}

Since IoT healthcare devices operate in an interconnected and interdependent environment, new threats constantly emerge. Moreover, as IoT healthcare devices typically use in an unattended environment, intruders may maliciously access these devices. Eavesdropping can access privately-owned information from the communication channel because IoT devices are usually linked through wireless networks. In addition to these security issues, IoT devices cannot afford to incorporate advanced security features because of their limited energy and processing power. Therefore, it is essential to study the effect of malicious attacks on the energy consumption of smart healthcare devices and show their impact, as smart healthcare systems are much more vulnerable and sensitive to their privacy and security. More importantly, considering the massive amount of smart healthcare devices on the market, the impact of energy consumption attacks cannot be neglected.
Our main contribution is studying the effect of a practical combination of F-APs and DDoS attacks on smart healthcare devices. The main purpose of choosing DDoS and F-APs attacks as their impact on IoT security is high, so many researchers are working on solutions~\cite{kumari2023comprehensive}. Thus, we target the energy consumption of smart devices. In the first step, we design a smart system to measure the current consumption of smart healthcare devices. Also, we build a testbed to monitor the smart devices and capture devices' status, e.g., \emph{On} or \emph{Off}, network traffic, and energy consumption. We identify several critical influential factors, particularly in communication protocols, Attack Rate (AR), payload size, and victim devices' ports status. We study the impact of these factors on the victim devices' resource constraints, such as energy consumption. In the second part of the contribution, the attack continues, and the attacker disconnects the smart devices from the local AP by sending DDoS attacks. At the same time, we study the effect of EC-DDoS on energy consumption by sending malicious attacks to affect the energy resources of smart devices. Also, we implement real-time energy monitoring on real smart devices to register the effect of the attack.
In the last part of the contribution, we designed the F-AP to force the smart devices to connect to it once it disconnected from the local AP by DDoS attack. We designed the F-AP to automatically send malicious attacks affecting smart devices' energy consumption. 

This paper gives a valuable understanding of the effect of DDoS and F-AP attacks on the energy consumption of smart healthcare devices by presenting a testbed and accurate energy consumption tests. Energy consumption attacks can destroy smart devices and impact patients' lives.

\subsection{Organization of the paper}    
\label{SUBSEC:OUTLINE}
This paper is organized as follows: Section~\ref{SEC:Background} reviews related work about the different types of attacks. Section~\ref{Sec:AttackingScenarios} explain the attack scenario and assumptions. The testbed scenario, data collection, F-APs setup, and the most important influential factors are presented in Section~\ref{SEC:Proposed}.
Section~\ref{SEC:Exp} describes different results regarding network scans, disconnections caused by DDoS attacks, energy consumption measurement, and the effect of EC-DDoS and F-APs on energy consumption. Some concluding remarks can be found in Section~\ref{SEC:CONCLUSIONS}.

\section{Related Work}
\label{SEC:Background}
\subsection{Fake Access Points Attacks}
\label{SUBSEC:FAPs}
One of the most challenging security problems for wireless networks is detecting F-AP attacks. This attack is also called the rogue AP attack or the evil twin attack~\cite{metwally2022detecting}. 

Detection of rogue AP attacks in the wireless network of a smart healthcare system is an essential aspect of wireless security~\cite{zhang2022sovereign}.
A rogue device detection system using various techniques such as site survey, noise checking, MAC address list checking, and wireless traffic analysis has been proposed in~\cite{Ibrahim2007}. The authors concentrated on detecting internal rogue devices, such as devices connected via a wireless network and used by employees on a corporate network. But this approach cannot be applied to IoT devices due to resource constraints.

Mehndi \emph{et al.}~\cite{Samra2015} proposed an approach that considers the Mac Address, Service Set Identifier (SSID), and signal strength of the AP to decide whether the AP is rogue or not. In detecting authorized APs, the MAC addresses of all visible APs are matched against a list of authorized APs. Tools such as 
Ettercap\footnote{https://ettercap.github.io/ettercap/},
Wireshark\footnote{http://www.wireshark.org/}, and Snort\footnote{http://www.snort.org/} are used for filtering instances where the MAC address is spoofed.
While Kilincer, Ertam, and Şengür~\cite{Kilincer2020} proposed an automated technique for detecting and preventing F-APs attacks in the network of IoT devices. The proposed experiment uses a Single Board Computer (SBC) and a wireless antenna (ODROID module). The operation was about:  
\begin{inparaenum}[1)]
\item creating an F-APs, \item scanning the surroundings using the SBC and WiFi modules, and \item detecting fake AP broadcasts. 
\end{inparaenum}
The F-APs have been assigned to an unauthorized Virtual Local Area Network (VLAN). This study \cite{Kilincer2020} is limited and focuses on F-APs attack detection and prevention. However, the data collected about the network and some attacks are still possible without connecting.

\begin{figure}[ht]
	\includegraphics[width=\linewidth]{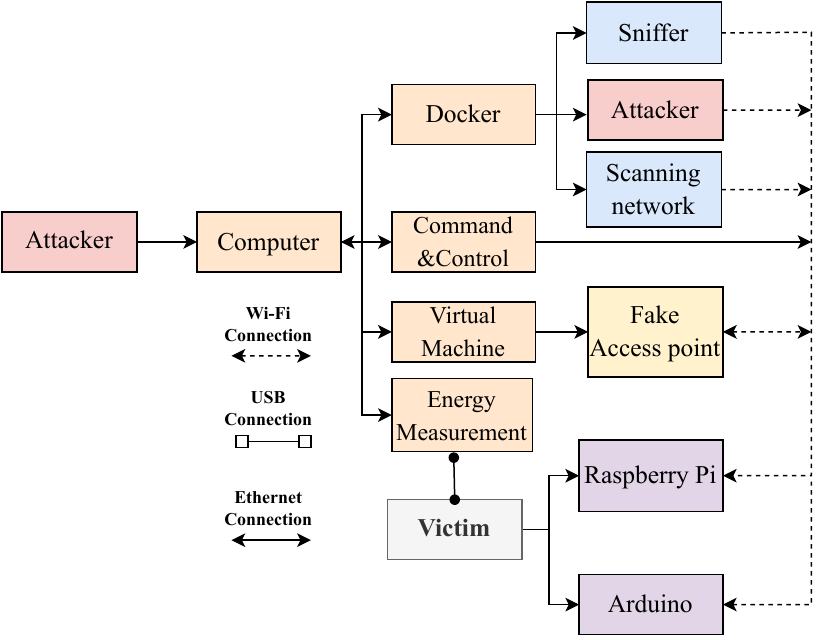}
	\centering
	\caption{Testing Environment.}
	\label{FIG:TE}
\end{figure}
\subsection{DDoS and EC-DDoS Attacks}
\label{SUBSEC:MDDoS}

DDoS and EC-DDoS attacks are security threats by attackers that enter the WiFi network coverage area and inject many different forged packets. Adversaries use this attack for two purposes, restricting usage of the WiFi bandwidth and preventing licensed users from communicating with the licensed AP to paralyze or reduce the WiFi network's performance~\cite{Zhanyong2017}.

Different studies have focused on the impact of DDoS attacks on Web servers when compromised IoT devices launch the attack. For example, Kambourakis~\cite{Kambourakis2017}, Marzano~\cite{Marzano2018}, Tushir~\cite{Tushir2020}, and Kolias~\cite{Kolias2017} discussed the outbreak of the Mirai botnet (and its variants), which compromised IoT devices to launch a DDoS attack against data centers. They claim that even naive techniques can be used to take control of such devices and create a massive and highly disruptive army of zombie devices.
Liu and Qiu~\cite{Chibiao2017} examined de-authentication and disassociation DDoS attacks, where the attacker overwhelms the wireless device through fake de-authentication and disassociation packets. They show that Transmission Control Protocol (TCP) throughput drops and User Datagram Protocol (UDP) packet loss rises by increasing the AR. They have developed a client-device-based queuing model to show that the current IEEE 802.11w standard cannot resolve de-authentication and disassociation at high attack rates.
Moreover, there are different approaches to monitoring IoT devices' energy consumption to detect IoT cyberattacks. Tushir \emph{et al.}~\cite{Tushir2020} quantitatively studied the impact of DDoS attacks on smart home IoT devices and their energy consumption. 
However, they did not present any detection or mitigation solutions.

Despite this, many developed methods exist to detect and prevent DDoS and F-AP attacks in IoT systems. In our approach, we mainly focused on the impact of the combination of DDoS, EC-DDoS, and F-APs attacks on energy consumption, response time, and connectivity of smart healthcare devices.

\section{Attack Scenarios and Assumptions}
\label{Sec:AttackingScenarios}
\begin{figure}[!ht]
    \centering
	\includegraphics[width=\linewidth]{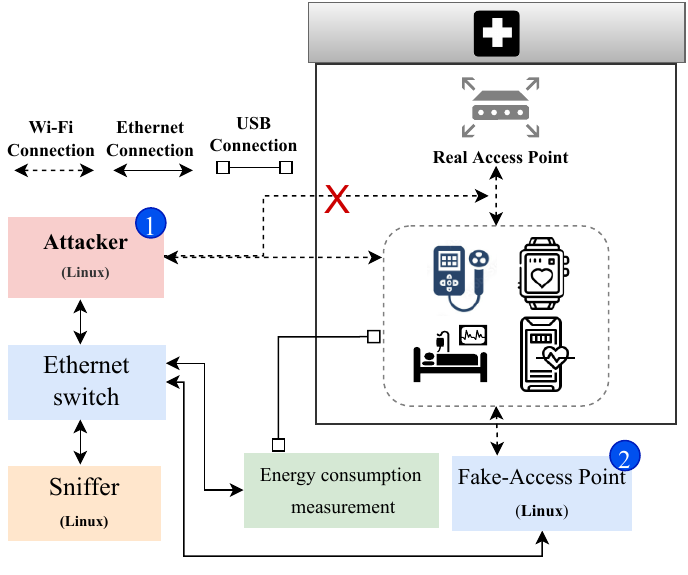}
	\caption{Attacking Scenarios.}
	\label{FIG:AttackSC}
\end{figure}

\emph{Attack Scenarios.} \textbf{Figure~\ref{FIG:AttackSC}} illustrates the potential scenarios where DDoS, EC-DDoS, and F-APs attacks can be applied. First, F-AP is designed to look like the actual AP. In those scenarios, the attacker can set a F-AP to launch different attacks to affect the energy resources of the smart healthcare devices. The F-AP signals could be more vital to the victim than the actual AP. Once disconnected from the actual AP by sending DDoS attacks, the tool forces smart healthcare devices to automatically reconnect to the F-AP, allowing the attacker to intercept all the traffic to that smart healthcare device, such as Man-In-The-Middle (MITM) attacks. 
Sniffing tools can also be applied to get or edit the information sent or received from the victim's devices. 
Additionally, the attackers may use F-AP and EC-DDoS attacks to destroy smart healthcare devices by affecting resource usage, e.g., energy consumption.

\emph{Assumptions.} The attacks we consider require the attacker to send DDoS attacks to force the device to disconnect from the legitimate AP and affect its resources. Then the adversary could set up the F-APs attack at different locations. The adversary may set the F-APs at a distance from the victim to avoid being caught. 
As a result, the smart device is connected to the F-APs created by the attacker.
The potential F-APs attack relies on sniffing over the WiFi to capture all packets traveling to and from the monitored smart device. Also, the F-AP is designed to affect the energy consumption of smart devices by automatically sending malicious attacks to connected smart devices.

Therefore, in this paper, we investigate the effect of F-APs and EC-DDoS attacks on smart healthcare devices' energy consumption by implementing them with malware designed to increase the energy consumption of smart healthcare devices and destroy them.

\begin{figure*}[ht]
	\includegraphics[width=\textwidth, height=7.5cm]{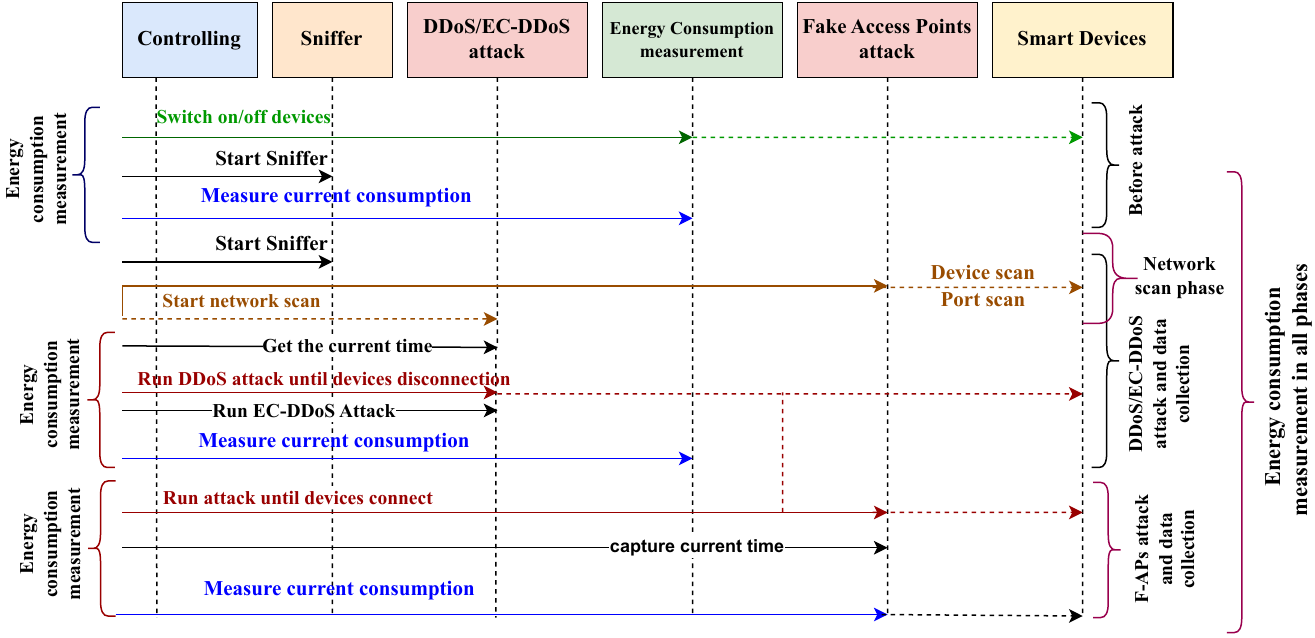}
	\centering
	\caption{Sequence diagram showing an attacker intercepting and affecting energy measurement of the smart healthcare devices. }
	\label{FIG:AttackSC1}
\end{figure*} 

\section{Proposed Testing Environment}
\label{SEC:Proposed}

This section investigates attack scenarios against smart healthcare devices and describes a testbed, network scan, and data collection process. The testbed is used to study the effects of DDoS and F-AP attacks on energy consumption and analyze devices and ports' status to identify their weak side points.

\subsection{Experiment Setup}
\label{SEC:Test}
There are two types of attacks on IoT devices: internal and external. Internal attacks happen when the adversary has access to the local network; this is possible by hacking the WiFi or gaining access to IoT devices. For example, attackers may gain access to the device by launching internal attacks to access a local Linux-based device remotely or by sending packets from outside the network for the external attack. For instance, if the attacker can force smart healthcare devices to disconnect from the actual AP and connect to the F-APs, then the adversary can send packets to the device outside the local network.

The testbed contains different smart healthcare devices, e.g., Arduino and Raspberry Pi, as proof of concept. Furthermore, three Linux-based images were created using Docker, as shown in Figure~\ref{FIG:TE}. These images involve \begin{inparaenum}[1)] \item an attacker sending malicious packets to the victim devices, \item a sniffer for capturing WiFi traffic, and \item a control system to scan the network and get different information about the port and devices' status. 
\end{inparaenum}

Moreover, we created the F-APs using a TP-Link TLWN772N USB adapter and a Linux-based software system. Then, we 
designed a smart meter to measure the energy consumption 
of smart healthcare devices using a non-invasive current sensor~\cite{Zainab2019} with an Arduino and some other resistors.
Also, we used different software tools for attacking data generation and collection. On the adversary side, we used Nmap\footnote{https://nmap.org/} to launch a network scan and identify devices' status, such as \emph{online} or \emph{offline}, IP address, and MAC address. Then, we ran a TCP/UDP port scan on the victim devices to identify port status (open/closed, filtered/not filtered, and others). 
Furthermore, we used hping3\footnote{https://www.kali.org/tools/hping3/} to generate DDoS and EC-DDoS attacks by adjusting the AR, source IP address, destination IP address, payload, attack type, flags of TCP sessions (SYN, ACK, FIN, push, or urgent), and port types. In addition, we used hostapd (host access point daemon) and dnsmasq with a TP-Link TLWN772N USB adapter to create the F-APs. Hostapd is a user-space daemon software enabling a network interface card to act as an AP and authentication server. Dnsmasq is a lightweight, easy-to-configure DNS forwarder designed to provide DNS (and, optionally DHCP and TFTP) services to a small-scale network. The TP-Link TLWN772N is a USB adapter that acts as a F-AP.  

We used tshark to evaluate the impact of EC-DDoS and F-APs attacks on the resource constraints of smart healthcare devices and capture WiFi traffic. 
Figures~\ref{FIG:AttackSC} and ~\ref{FIG:TestBed} show the attacking scenarios.
Different stages are used to run our experiment. In the first stage, we measure the energy consumption
once the device is turned \emph{On}. 
Another measurement is used when the device has connected to the AP. Then, we run a network scan to capture the port and device status. Once we ensure that the device is connected to the Internet, we send DDoS and EC-DDoS attacks for two purposes: first, to consume more energy, and second, to disconnect the device from the local AP. Then, we run energy consumption measurements to calculate the energy consumption of the devices and study the devices' behaviors under DDoS and EC-DDoS attacks. Next, we run the F-APs attack on smart devices. We first check whether the devices are already disconnected from the local AP. We then force the smart devices to connect to the F-APs and finally start measuring the energy consumption of smart healthcare devices. Also, we study the behavior of smart devices in terms of energy consumption and connectivity. The F-AP works as a MITM attack. The F-APs used for different purposes: \begin{inparaenum}[1)] \item monitoring the devices, \item sending malicious packets to consume more energy, and \item affecting the CPU usage of the smart healthcare devices. \end{inparaenum} 

\begin{algorithm}
\caption{the affecting of F-APs and DDoS attacks on energy consumption}
\begin{small}
\begin{algorithmic}[1]
\Procedure{SmartDevice}{$a$} \Comment{\textit{consume more energy of  (a).}}
    \State Sniff air for network scanning
    \State E1:energy consumption before attack
    \State E2:energy consumption after attack
    \If{$a = connected$} \Comment{\textit{a is connected to the AP}}
        \State Calculate energy consumption
        \State Disconnect using DDoS attack
    \ElsIf{$a =disconnected$} \Comment{\textit{from the actual AP}}
        \State Send DDoS attack
        \State Calculate energy consumption after attack
    \EndIf
    \If{$E1($a$) < E2($a$)$}
        \State Connect the device to F-APs
        \State Send malicious attack to consume more energy
    \ElsIf{$E1($a$) < E2($a$)$} \Comment{\textit{energy consumption after attack}}
        \State Send another packets of DDoS attack
        \State Consume more energy and disconnect the devices
    \Else
        \State Try to consume more energy
        \State Calculate energy consumption, AR, survival duration (SD), and threshold(AR)
    \EndIf
    \While{$E1 \not= 0$}  \Comment{\textit{Consume more energy to destroy the device}}
        \State $E1($a$) \leftarrow E2($a$)$  
    \EndWhile  
\EndProcedure
\end{algorithmic}
\end{small}
\end{algorithm}

\subsection{Collecting Data}
\label{SUBSEC:DataCollection}
\begin{figure}[ht]
	\includegraphics[width=\linewidth]{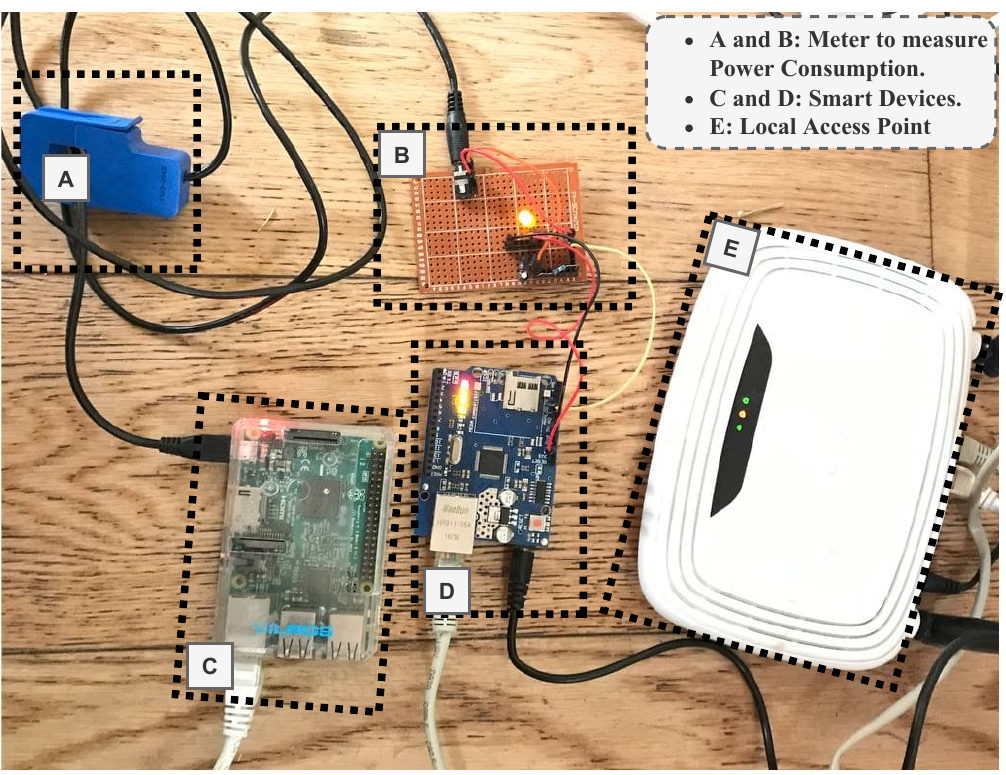}
	\centering
	\caption{Proof of Concept for Wireless Network smart healthcare devices.}
	\label{FIG:TestBed}
\end{figure}

We built a smart healthcare test environment by deploying different smart devices, as shown in Figure~\ref{FIG:TestBed}. Data is aggregated from various smart devices for analysis purposes. Moreover, the aggregated data is categorized into behavioral and network data. Behavioral data refers to the status of the smart devices, like \emph{On} or \emph{Off}, and the device's readings. Network data refers to smart healthcare devices' TCP and UDP packet data. We integrate this data to learn typical behaviors in a smart healthcare environment.

\begin{figure}[ht]
	\includegraphics[width=\columnwidth]{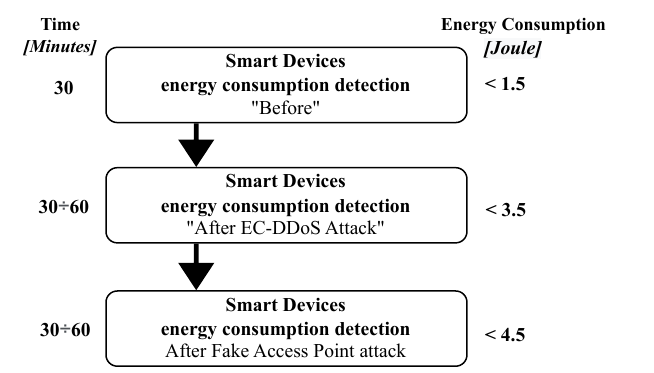}
	\centering
	\caption{Energy Consumption affects before and after attacking smart healthcare devices. }
	\label{FIG:PowerConsumption}
\end{figure}

Figure~\ref{FIG:AttackSC1} shows different phases of our attack scenario; in the first phase, we control the smart devices by switching them \emph{On} or \emph{Off}; this is essential to calculate the energy consumption of the smart devices before launching any attack. In the second phase, we measure the energy consumption of the smart devices for at least $30$ minutes to calculate the average energy consumption before launching any attack; then, we start sniffing the network to get information about ports and devices' status. In the final phase, we launch DDoS and EC-DDoS attacks to impact smart devices' connectivity and energy consumption. Simultaneously, the energy consumption of appliances is measured to show the impact of this attack on the energy consumption of smart healthcare devices. After that, once the devices are disconnected from the legitimate AP using DDoS attacks, we calculate AR, Survival Duration (SD), and the threshold of the AR.
Next, we force the devices to connect to the F-APs, where the fourth phase will start. We start monitoring and collecting information about the smart devices using F-APs facilities in this phase.
We then launch malicious attacks through the F-APs to consume more energy and study the behaviors of smart devices' energy consumption under F-APs attacks. 
Through that, we achieve the main purpose of destroying smart healthcare devices by using energy consumption attacks caused by F-AP and EC-DDoS attacks as shown in Figure~\ref{FIG:PowerConsumption}.

\subsection{Setting up a Fake Access Point}
\label{SUBSEC:FAP}

We used hostapd (host access point daemon) and dnsmasq with a TP-Link TLWN772N USB adapter to create the F-APs attack, broadcast a fake signal, and capture the victim's packets. The F-AP configures the same SSID, Basic Service Set Identifier (BSSID), broadcast channel, and security settings as the legitimate AP. The scenario with the F-APs attack is reported in Figure~\ref{FIG:AttackSC}.
The monitor mode of the F-APs is enabled using \emph{airmon-ng start} for capturing attack injections or packets from and to the smart healthcare devices. The F-APs work as a MITM attack to capture packets transferred between the smart devices and the server. Moreover, the F-AP is designed to send malicious attacks to consume more energy of the connected smart healthcare devices\footnote{https://github.com/developerZA/EnergyConsumptionAttack.git}.

\subsection{Determining the weak side}
\label{SUBSEC:influentialFactor}

In this section, we focus mainly on the impact of disconnections and power consumption, considering victim devices that differ in their hardware, e.g., CPU, WiFi chip, and memory. Therefore they respond differently to a given attack. Arduino and Raspberry Pi are considered.

During an attack, the port state and communication protocol can significantly influence victims’ responses. Accordingly, we launch three attacks: TCP-SYN, UDP, and Internet Control Message Protocol (ICMP) echo request attacks. In the case of UDP and TCP-SYN attacks, we launch attacks against ports with different states, e.g., open, closed, filtered, and open-filtered. With ICMP attacks, do not specify the port number.
Moreover, being smart IoT devices resource-constrained, their behavior may differ dramatically according to the payload size of attack packets. As a result, we set the payload of attack packets to $0$~B, i.e., no payload (NP), and $1500$~B, i.e., high payload (HP).
Although it has been recognized that the impact of DDoS, EC-DDoS attacks are directly related to AR. Section~\ref{SUBSEC:Disconnection} analysis the impact of the AR on service disruption by calculating the minimum AR that causes the devices to be disconnected from the legitimate AP. The impact of EC-DDoS and F-APs attacks on the energy consumption of smart healthcare devices has also been studied in Section~\ref{SEC:Exp}.

\section{Experimental results and analysis}
\label{SEC:Exp}

In this section, we describe an experimental workplace to test the effect of DDoS, EC-DDoS, and F-APs attacks on the energy consumption of smart healthcare devices. This experiment focuses on collecting incoming malicious attacks and the usage statistics of a victim device and analyzing the attack effects on the victim devices in terms of energy. The network scan of the smart devices is used to obtain the status of the ports and then to determine the weak side of smart devices by calculating their AR and SD. Moreover, we study the effect of EC-DDoS and F-APs attacks on energy consumption.

\subsection{Network Scan}
\label{SUBSEC:NETWORKSCAN}
The network scan gathers information about the victim's smart devices, such as \emph{online} or \emph{offline} status, IP address, and MAC address. The port scan permits an attacker to discover the status of TCP and UDP ports. The possible states of the used ports are: open, closed, filtered, and open-filtered. 
Table~\ref{tab:networkscan} reports the status of ports for the devices used in our testbed.
\begin{table}[htbp]
\renewcommand{\arraystretch}{1.7}
	\centering
	\caption{Network scan result in terms of port status for TCP and UDP protocols.}.
	\label{tab:networkscan}
    \setlength{\leftmargini}{10pt}
    \begin{footnotesize}
    \begin{tabular}{ |m{0.2\textwidth}<{\centering}  |m{0.35\textwidth}<{\centering}  
    |m{0.35\textwidth}<{\centering}
    |m{0.35\textwidth}<{\centering}
   |}
        \hline
        \textbf{Device} & \textbf{TCP scanned ports} & \textbf{ UDP scanned ports}\\
        \hline
        Raspberry Pi & 
                    3 open, 998 open-filtered, 65389 filtered and 0 closed ports & 4 open and 700 open-filtered and 0 closed ports  \\
        \hline
       	Arduino & 1 open, 22 filtered, 1000 open-filtered and 0 closed ports & 1000 open-filtered ports \\
        \hline
    \end{tabular}
    \end{footnotesize}
\end{table}
\subsection{Attack Rate and DDoS Attacks}
\label{SUBSEC:Disconnection}

We consider threshold AR as the minimum AR measured in
Packets Per Second (PPS) that disconnects the victim device from the AP. The SD is the time duration between the start of an attack and the device disconnection caused by the attack. We set the maximum attack duration between $8$ and $30$ minutes. We have launched ICMP and TCP-SYN/UDP attacks on the victim devices' open, filtered, and closed ports to collect their threshold AR and SD. 
The AR applied to the Raspberry Pi is between $500$ to $10,000$ PPS for both NP and PH. In contrast, the AR sent to the Arduino for NP attacks is between $100$ to $800$ PPS, as the threshold AR is $800$ PPS. We did not use a PH attack against the Arduino because it disconnects with minimal AR.
Table~\ref{tab:SD} reports the average SD in minutes for the smart devices. We can see that the Arduino device disconnects in all cases with different attacks. Instead, the Raspberry Pi disconnects only with low packets. 
\begin{table}[!ht]
\renewcommand{\arraystretch}{1.7}
    \centering
	\caption{Survival Duration (SD) caused by DDoS attack.}
	\renewcommand{\arraystretch}{1.2}
	\label{tab:SD}
    \setlength{\leftmargini}{10pt}
    \begin{footnotesize}
\begin{tabular}{|l|cc|cc|}
\hline
\multicolumn{1}{|c|}{\multirow{2}{*}{\textbf{\begin{tabular}[c]{@{}c@{}}Survival \\ Duration\end{tabular}}}} & \multicolumn{2}{c|}{\textbf{Raspberry Pi}}                                 & \multicolumn{2}{c|}{\textbf{Arduino}}                                      \\ \cline{2-5} 
\multicolumn{1}{|c|}{}                                                                                       & \multicolumn{1}{c|}{\textit{NP {[}Minutes{]}}} & \textit{PH {[}Minutes{]}} & \multicolumn{1}{c|}{\textit{NP {[}Minutes{]}}} & \textit{PH {[}Minutes{]}} \\ \hline
SD (ICMP)                                                                                                    & \multicolumn{1}{c|}{7.58}                      & none                      & \multicolumn{1}{c|}{3.6}                       & 3.13                      \\ \hline
SD (TCP)                                                                                                     & \multicolumn{1}{c|}{6.2}                       & none                      & \multicolumn{1}{c|}{3.3}                       & 2.44                      \\ \hline
SD (UDP)                                                                                                     & \multicolumn{1}{c|}{7.8}                       & none                      & \multicolumn{1}{c|}{3.8}                       & 2.44                      \\ \hline
\end{tabular}
\end{footnotesize}
\end{table}
The Raspberry Pi survives with a higher AR than the Arduino, with $20$~k packets at NP. The AR of the Arduino is $800$ packets at NP and $200$ packets at PH. Looking at the tshark files, we can see that the Raspberry Pi broadcasts probe requests and sends de-authentication packets to the legitimate AP. The main difference between the smart devices is that the Raspberry Pi has more powerful hardware than the Arduino.

\begin{figure}[ht]
	\includegraphics[width=0.9\columnwidth]{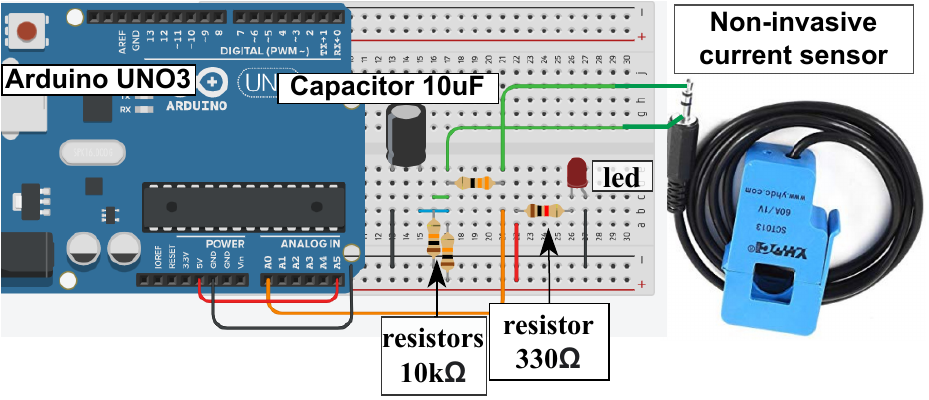}
	\centering
	\caption{Circuit for measuring current consumption.}
	\label{FIG:Circuit}
\end{figure}

Through the experiment, when we calculate the received AR by the victim devices, we find that the victim devices rarely receive the actual AR sent by the attacker. For example, we sent about $15$~k packets to the open ports of the Raspberry Pi; the received packets were about $14544$ packets. 
Also, we can notice that the increase in the average packet rates sent by the attacker causes an approximately logarithmic increase in the received packets by the victim.

\subsection{Energy Measurement and EC-DDoS Attacks}
\label{SUBSEC:ENERGYCONSUMPTIONMEASURMENT}

We developed a smart circuit using a non-invasive current sensor, as shown in Figure~\ref{FIG:Circuit} to measure the current consumption of smart healthcare devices. This smart circuit samples voltage, ampere, watt, and current per second.
The current consumption values for each smart healthcare device are stored in the database (DB). 
In our experiment, we use the Joule (J) values to calculate the energy consumption of smart devices. 

To calculate the energy consumption of the devices versus the incoming attack reception rate of the victim devices, we need to collect the data from both the sensors and the tshark data. Therefore, all data relevant to this experiment is stored automatically in the DB.
\begin{figure}[!ht]
	\includegraphics[width=0.9\columnwidth]{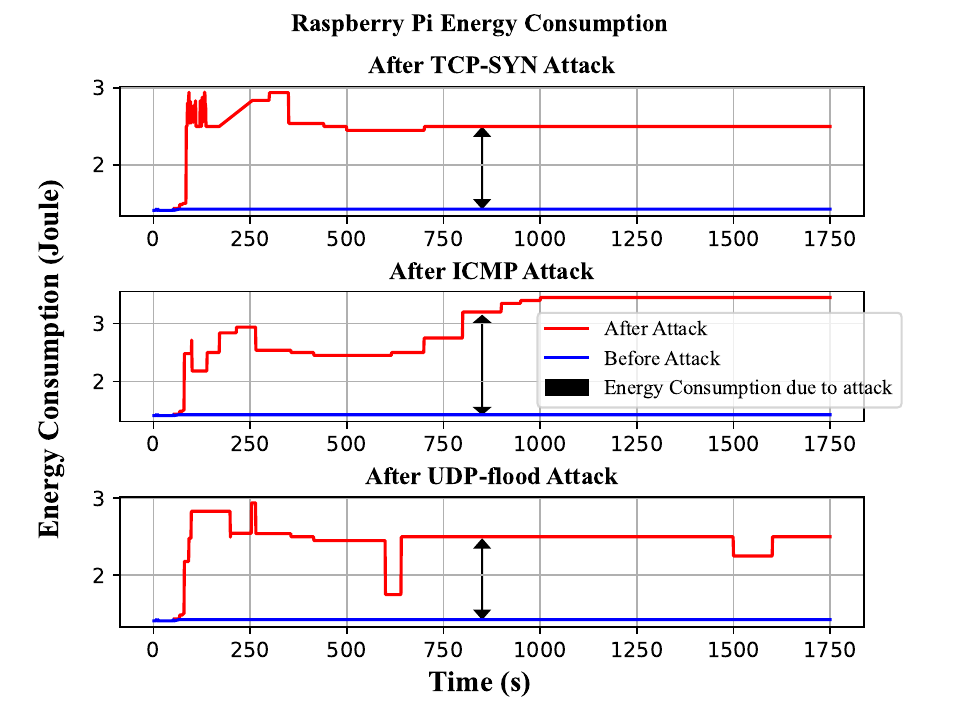}
	\centering
\caption{Raspberry Pi Energy Consumption under EC-DDoS Attack.}
	\label{FIG:EnergyConsumptionRP}
\end{figure}
During packet collection, the attacks are sent using the same TCP, UDP, and ICMP flood commands. Using the topology
depicted by Figure~\ref{FIG:AttackSC}, the malicious TCP, UDP, and ICMP traffic are separately sent to the victim device, while all usage statistics in terms of energy consumption and connectivity are recorded on the victim device. Each attack is simulated for a duration of $1$ second for a total of $30$ minutes, and all usage statistics are recorded for the same duration.

Figure~\ref{FIG:EnergyConsumptionRP} shows the device's energy consumption when its status is \emph{On} in the absence of attacks on that device. The standard energy consumption of the Raspberry Pi is between $1.410$~J and $1.420$~J per second. However, the current consumption varies from $1.410$~J to more than $3.3$~J per second after launching TCP-SYN attacks on open ports of the Raspberry Pi.
In contrast, we can notice that the energy consumption increases to more than $3.60$~J per second after launching ICMP attacks on open ports of the Raspberry Pi. 
Additionally, the energy consumption fluctuates between $1.4$~J and $3.50$~J per second after launching a UDP flood attack because of the overload that might have happened on the Raspberry Pi's open ports.

\begin{figure}[!ht]
	\includegraphics[width=0.9\columnwidth]{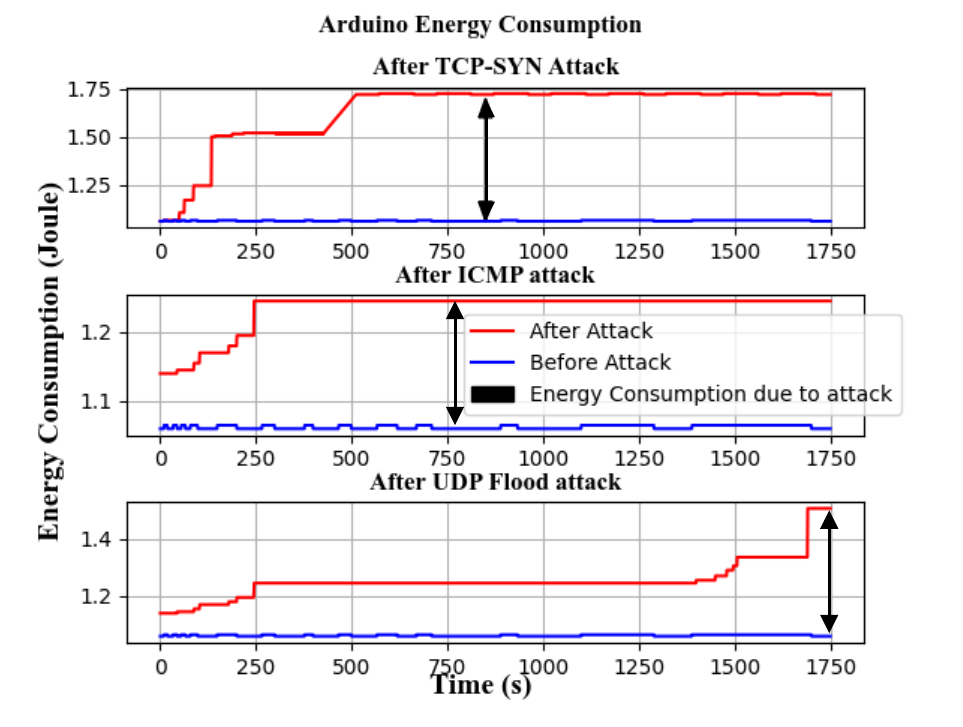}
	\centering
	\caption{Arduino Energy Consumption under EC-DDoS Attacks.}
	\label{FIG:EnergyConsumptionAR1}
\end{figure}

Figure~\ref{FIG:EnergyConsumptionAR1} shows the current consumption of the Arduino when its status is \emph{On}
in the absence of attacks. The standard energy consumption of the Arduino is between $1.060$~J and $1.065$~J per second. 
In contrast, the energy consumption varies from $1.065$~J to more than $1.75$~J per second after launching TCP-SYN attacks on NP. At the same time, the energy consumption increases slightly from $1.15$~J to $1.25$~J per second after sending an ICMP attack. The UDP flood attack causes an increase in energy consumption from $1.25$~J to more than $1.50$~J per second.

Note that the EC-DDoS attack rates sent by the attacker are below the threshold of DDoS attack rates that cause disconnection on smart healthcare devices.
In the next section, we study the smart devices' energy consumption behavior under F-APs attacks.  
\begin{algorithm}
\caption{Fake Access Points Attack}
\begin{small}
\begin{algorithmic}[1]
\Procedure{SmartDevice, F-APs}{$a$,$b$} \Comment{\textit{consume more energy of  (a).}}
    \State Sniff air for network scanning
    \State Measure energy consumption of SH
    \If{$a \subseteq b$}
        \If{$a = connected$} \Comment{\textit{a is connected to F-APs}}
            \State Sniff air for network scanning
            \State Send malicious packets
            \State Calculate energy consumption after attack
        \ElsIf{$a = NotConnected$} 
            \State Try to reconnect it to F-APs
        \EndIf
    \EndIf
    \While{$a \not\subseteq  b$}  \Comment{\textit{if there are no new devices}}
        \State Sniff air for finding new devices  
    \EndWhile  
\EndProcedure
\end{algorithmic}
\end{small}
\end{algorithm}

\subsection{Energy Consumption and F-APs Attacks}
\label{SUBSEC:F-APEnergy}
Once the devices are disconnected from the legitimate AP, the F-APs attack takes over its responsibility to consume more energy and monitor them.

The signal of the F-APs is more vital to the victim's smart devices than the legitimate AP. When the devices are disconnected, the signal from the F-APs will be sent to the smart devices to force them to connect to affect their energy resources. Afterward, the monitoring mode of the F-APs will be enabled to monitor packets transferred from and to the smart devices. At this stage, the sniffer is essential to launch further attacks on the target device and collect information about it, such as IP and port status. The F-AP is designed to be more flexible in sending malicious packets automatically to affect the energy resources once the smart devices are connected.

The required time for the Raspberry Pi to connect to the F-APs is between $3$ and $5$ minutes. While the Arduino takes $7$ to $10$ minutes, sometimes we force the Arduino to connect to the F-APs.
\begin{figure}[!ht]
	\includegraphics[width=0.9\columnwidth]{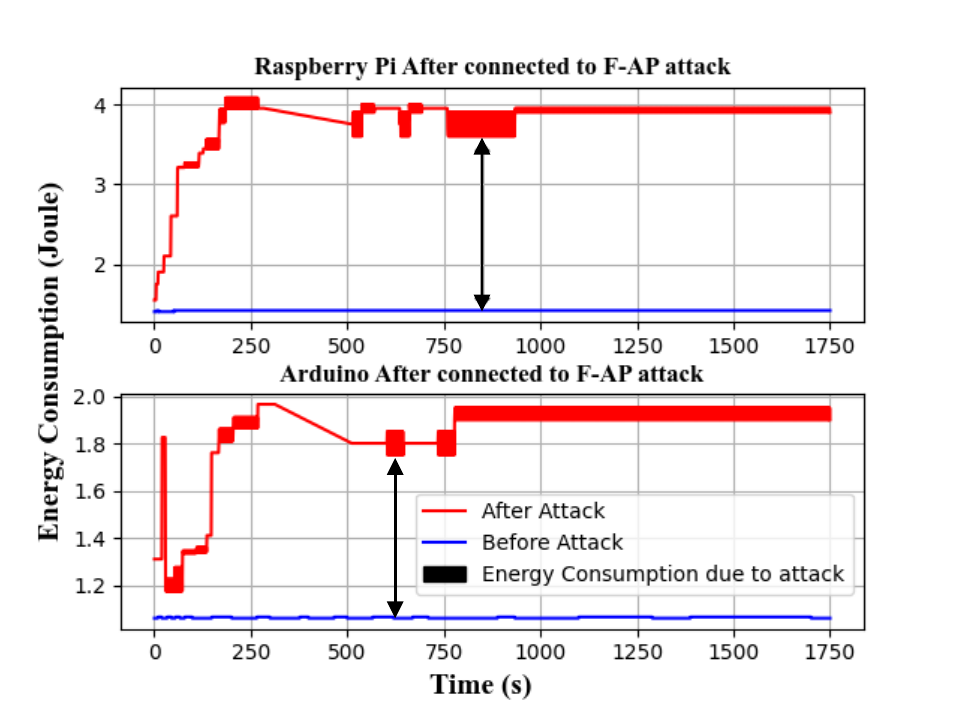}
	\centering
	\caption{Raspberry Pi and Arduino Energy Consumption under F-APs Attack.}
	\label{FIG:EnergyConsumptionF-RP}
\end{figure}
Figure~\ref{FIG:EnergyConsumptionF-RP} shows how the energy consumption of the Raspberry Pi changed after connecting it to the F-APs; the malicious packets were randomly selected and sent to the Raspberry Pi. The energy consumption increases to more than $4.00$~J per second. 
At the same time, the energy consumption of the Arduino increases slightly to reach more than $2.00$~J per second after connecting it to the F-AP. Therefore, we can conclude that the F-APs attack successfully affects smart healthcare devices' energy consumption.

\subsection{Results and Analysis}
\label{SUBSEC:ResAna}
In our experiment, we studied the effect of DDoS, EC-DDoS, and F-APs attacks against the Raspberry Pi and Arduino for about $30$ to $60$ minutes and measured the energy consumption. During such attacks, the smart devices continuously receive the packets and spend resources processing these packets.
\begin{figure}[ht]
	\includegraphics[width=0.9\columnwidth]{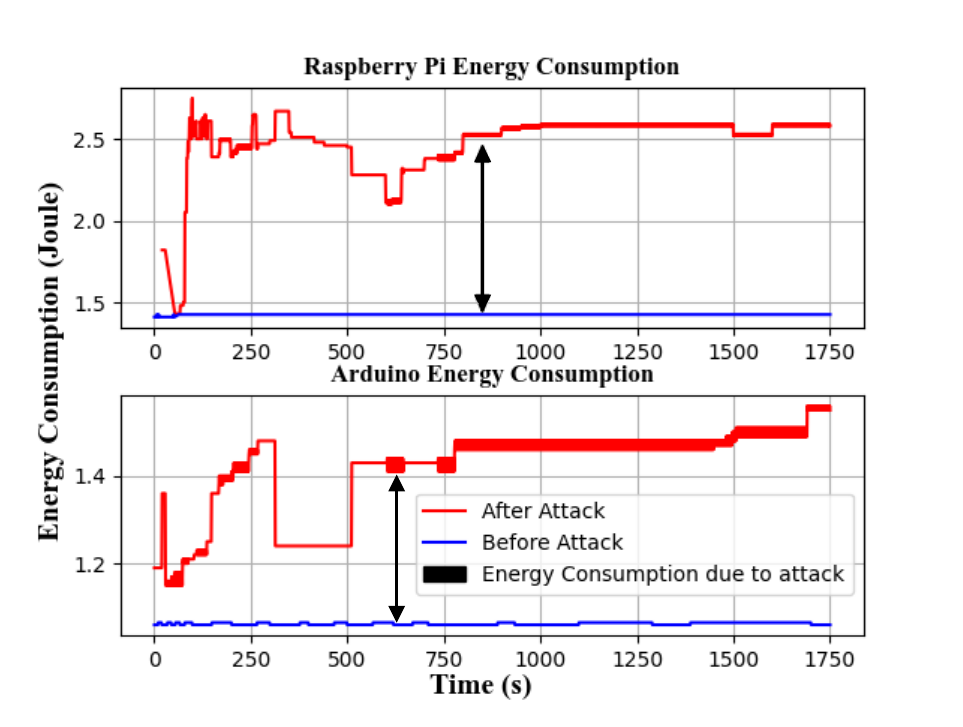}
	\centering
	\caption{Raspberry Pi and Arduino Energy Consumption under Attacks where the F-APs affect $45$\% of the energy consumption of the Raspberry Pi and the Arduino, while the affection of EC-DDoS attack is about $55$\%.}
	\label{FIG:EnergyConsumptionGeneral}
\end{figure}
Our analysis shows that effective DDoS attacks can be launched at NP if the victim replies to ICMP packets. ICMP or TCP-SYN/UDP attacks could be used on open and closed ports. However, to launch EC-DDoS attacks that cost the victim device’s maximum energy without being disconnected from the legitimate AP, the attacker can launch a PH TCP-SYN attack against open ports or ICMP attacks if the device responds to ICMP packets.

Moreover, to force the smart healthcare devices to connect to the F-AP attack, the signals of the latter should appear stronger to the victim than the legitimate APs. The attacker launches malicious attacks through the F-AP to induce maximal energy consumption without being disconnected by considering the threshold of the AR. Figure~\ref{FIG:EnergyConsumptionGeneral} shows the overall infection of EC-DDoS and F-APs attacks on both devices (Arduino and Raspberry Pi); as it can be seen, the energy consumption of the Raspberry Pi device varies from $1.42$~J to be more than $3$~J per second. At the same time, the energy consumption of the Arduino varies from $1.06$~J per second to more than $2$~J per second. 
It is observed that DDoS, EC-DDoS, and F-APs attacks significantly impact the energy consumption of IoT devices. When an IoT device is flooded with TCP, UDP, and ICMP packets, there are significant increases in energy usage, which might destroy the IoT devices in the end. This study offers a better understanding of energy consumption attacks caused by the combination of F-AP, DDoS, and EC-DDoS attacks on the smart healthcare system.
The analysis of such resource consumption will benefit the
deep understanding of DDoS and F-APs attacks’ impact on resource-constrained smart healthcare environments and facilitated future research on lightweight defense mechanisms against such attacks. 

%
%
\section{Conclusions and Future Work }
\label{SEC:CONCLUSIONS}
This paper studied the impact of EC-DDoS and F-APs attacks on the resource usage of different smart healthcare devices and, more specifically, on energy consumption. We first used Docker images to collect data, scan the smart devices' networks, and sniff the network.
Then, we calculated the AR, SD, and threshold of the AR on the victim side. The main purpose of the calculation is to study the effect of DDoS attacks on the connectivity of smart healthcare devices. We also studied other influential factors such as ports, device state, attack type (i.e., protocols used), and AR. We then analyzed the impact of DDoS, EC-DDoS, and F-APs attacks on the energy consumption of smart devices. Specifically, we designed the F-APs attack to affect the energy resources of the smart devices by automatically sending malicious attacks to the connected smart healthcare devices. Through our work, we offer a better understanding of the effect of DDoS, EC-DDoS, and F-APs attacks on smart healthcare devices' energy consumption and connectivity within a wireless network. In the future, we will also study the effect of the combination of DDoS attacks and F-APs on the memory usage of smart healthcare devices. In addition, we will look for appropriate tools and strategies that consider the resources of the smart devices to protect the smart healthcare system from energy consumption attacks.

\appendix

\bibliographystyle{splncs04}
\bibliography{main}
\end{document}